\documentstyle[12pt]{article}
 \begin{document} 
 \addtolength{\topmargin}{-1cm}
 \title{ Theoretical Study of Small 
          (NaI)$_n$ Clusters
          } 
 \author{
  A. Aguado, A. Ayuela, J.M. L\' opez and J.A. Alonso   \\ \normalsize
  Departamento de F\'\i sica Te\'orica, Facultad de Ciencias,
  \\ \normalsize Universidad
  de Valladolid, 47011 Valladolid, Spain. \\ 
  }
%\date{}
 \maketitle
 \begin{abstract}
 
 A systematic theoretical study of stoichiometric clusters
$(NaI)_n$ up to $n=15$ is performed
 using the {\em ab initio} Perturbed-Ion (PI) model.
The structures obtained are compared to previous pair potential
results, and observed differences between $(NaI)_n$ clusters and previous
{\em ab initio} results for other alkali halide clusters are discussed.
$(NaI)_n$ clusters with $n$ up to $15$ 
do not show yet a marked preference for geometries which are
fragments of the bulk lattice.
Instead, stacks of hexagonal rings or more open structures are
obtained as ground structures in clusters with $n=3$, $6$, $7$, $9$, $10$, $12$, $13$ and $15$, indicating that convergence to bulk structure is not
achieved yet at this size range. 
Low lying isomers which are fragments of the crystal lattice exist,
nevertheless, for those cases. The binding energies show that clusters
with $n=(4)$, $6$, $9$ and $12$ molecules are specially stable. 
The binding energy has been decomposed in contributions which allow for an
intuitive interpretation. Some electronic 
properties like ionization potentials and electronic
energy levels are also studied.
\end {abstract}

\bigskip

 

\newpage

\baselineskip 20pt
\section{Introduction}

The structural and electronic properties of a material depend primarily
on the state of aggregation\cite{Ash76}. 
Small clusters often present significant physical
and chemical differences with respect to the molecule and the bulk material.
Cluster studies can provide valuable insight on the development of the
properties of condensed matter from those of isolated atoms or molecules.
One important source of information about cluster properties is provided by 
the abundance patterns emerging from the mass spectra. The relative 
intensities of cluster peaks are interpreted as being indicative of variations
in the cluster stability with size.
The measured population usually shows large 
drops after 
some particular sizes (often the population is also a maximum at those
sizes) and those particularly stable clusters are known as "magic clusters".
The magic numbers have been explained by using either electronic or 
structural       
models. For example, Van der Waals clusters $X_n$  appear to
 form compact icosahedral 
structures\cite{Ech81}
which exhibit enhanced stabilities for completed layers at values of 
n=13,55,147,... 
. In contrast, the
stability variations of alkali metal clusters have been interpreted in terms of
the electronic-shell model\cite{Hee87}.
A direct experimental determination of the spatial 
distribution of atoms or molecules in a cluster is a very difficult task;
however the knowledge of the cluster structure is a 
prerequisite for a quantitative 
discussion of many properties. 
Drift tube studies, which measure the mobility of a cluster ion through an
inert buffer gas under the influence of a weak electric field, provide
valuable information about the cluster geometries\cite{Hel91,Jar95}.
A theoretical study of the most stable structures
and their evolution as a function of the number of atoms or molecules in the
cluster is thus of the upmost importance for the interpretation of the
mobility experiments.

There are several experimental techniques for the production
of small clusters of alkali-halide materials:
particle sputtering\cite{Cam81}, inert-gas condensation\cite{Pfl85}
and laser vaporization\cite{Con88,Twu90}.
In this paper we are interested in
stoichiometric $(NaI)_n$ clusters.
Large non-stoichiometric ``ionized'' $(NaI)_nNa^+$ clusters have been produced
and their mass spectrum analyzed by Martin and coworkers\cite{Mar96}.
The spectrum is characterized by a periodic structure which has been
interpreted as revealing a definite pattern of growth: those clusters grow as
cuboids (clusters with rectangular faces and $n_1$ x $n_2$ x $n_3$ atoms, 
where $n_1$, $n_2$ and $n_3$ are the number of atoms along the three edges), 
and the observed intervals between minima in the spectrum
correspond to the number atoms
necessary to cover one face of the cuboid. The mass spectrometry techniques
detect only ionized clusters, although one can expect that the formation
pattern of neutral $(NaI)_n$ clusters will be analogous (neutral
stoichiometric $n_1$ x $n_2$ x $n_3$ cuboids are, however, not possible if
$n_1$, $n_2$ and $n_3$ are all odd numbers). Kreisle and
coworkers\cite{Mai96} have performed mobility measurements for
$(NaI)_nNa^+$ clusters up to $n=18$.

Dieffenbach and Martin have calculated the lowest energy 
structures of $NaI$ clusters using pair 
potential models\cite{Die85}. In pair potential
studies for $NaCl$ clusters, Phillips et al. \cite{Phi91} found
very small differences in binding energy between isomeric configurations
of the same cluster.
 In order to check if the approximations inherent in a
pair potential formulation have influence on the predicted geometries we use the
{\em ab initio} Perturbed Ion (PI) model, in which the radial 
relaxation of the wave functions
of the $Na^+$ and $I^-$ ions due to the interaction with the 
environment is taken into account.
Here we study
$(NaI)_n$ clusters up to $n=15$.
To our knowledge, these are the first ab initio
calculations on such clusters.
The PI model, which has been applied to study $(NaCl)_n$ clusters
in previous works\cite{Ayu92a,Ayu92b,Ayu95},
had been originally developed 
for the study of ionic materials in the crystal phase\cite{Lua90}.
The model is based
on the Theory of Electronic Separability (TES)\cite{Huz71,Lua87}
 and the restricted Hartree-Fock approximation.
The 
quantum-mechanical
interactions between the ions are treated in the following way:  
since the model assumes from the start that the cluster is built from
closed shell ions, when those ions
approach and the overlap becomes important,
the radial deformation of the charge clouds is taken into account
in a self-consistent way. The directional 
covalent effects are
considered less important than the radial deformations 
of the closed shell ions, and
this restricts the class of
 systems that can be treated.

It is also interesting
to compare our results with the results 
for $NaCl$ clusters\cite{Ayu95}, previously studied by the
PI model.
We want to 
see the influence of the ion size:
in the present case the size difference between the $I^-$ and $Na^+$ ions is
much larger than for $NaCl$ clusters.
The paper is organized as follows.
The PI model is presented in section II.
The results for  
the structures and binding
energies of $(NaI)_n$ clusters up to n= 15 are discussed in section III,
and
the conclusions are summarized in section IV. 

\section{Perturbed Ion model for clusters}

According to the Theory of Electronic Separability (TES)\cite{Huz71,Lua87},
when the
system under study is composed of weakly interacting groups, its wave function
can be expressed as an antisymmetrized product of group wave functions.
If these satisfy strong-orthogonality conditions\cite{Lyk56,Par56}, the
total energy is the sum of intragroup, or net, energies and 
intergroup interaction energies. 
For a cluster with a fixed three-dimensional arrangement  of the
ions, we consider each ion as a different group. 
All contributions to the total cluster energy
from a given group, the active ion A, can be collected in the effective 
energy of this group 
\begin{equation}
E_{eff}^A=E_{net}^{A}+ \sum_{S(\not= A)} E_{int}^{AS}=E_{net}^A+E_{int}^A.
\end{equation}
Here $E^A_{net}$ is the intragroup energy and $E^A_{int}$ is the
energy due to the interaction of the ion A with all the other ions.
Group wave functions can be
obtained by minimizing their effective energies if strong orthogonality
conditions are satisfied among the groups. 
The sum of the effective energies
is not equal to the total energy of the system
because this sum counts twice the intergroup energies. 
However, we can define an additive energy of the ion A by:
\begin{equation}
 E_{add}^A=E_{net}^A+\frac{1}{2}\sum_{S(\not= A)}E_{int}^{AS}=E_{net}^A+
\frac{1}{2}E_{int}^A
\end{equation}
and then the total energy of the system can be written 
\begin{equation}
 E=\sum_{A=1}^N E_{add}^A.
\end{equation}

The effective energies
can be expressed as  expectation values of appropriate effective Hamiltonians.
Each ion of interest (the active ion
A) will be a group in the sense of TES.
Its wave function  can be described by a single Slater determinant and
this can be obtained by
minimizing the expectation value
\begin{equation}
 E_{eff}^A(elec)=<\psi_A \mid H_{eff}^A\mid \psi_A>
\end{equation}
of the effective electronic hamiltonian\cite{Huz71}
\begin{equation}
 H_{eff}^A=\sum_{i=1}^{N_A} h_{eff}^A (i)+\sum_{1\le j< i\le N_A} r_{ij}^{-1},
\end{equation}
\begin{equation}
 h_{eff}^A(i)=T(i)-Z^A r_{iA}^{-1} +\sum_{S(\not=A)} [V_{eff}^S (i) + P^S (i)],
\end{equation}
where i, j run over the $N_A$ electrons of ion A and S counts the rest of the
ions of the cluster.
T is the kinetic energy operator and $Z^A$ is the nuclear charge of ion A.
$V_{eff}^S(i)$ represents the potential energy
of the ith electron (of ion A) in the field created by ion S
\begin{equation}
 V_{eff}^S (i) = - Z^S r_{iS}^{-1} + V_C^S (i) + V_X^S (i).
\end{equation}
The different parts of $V_{eff}^S$ are the electron-nucleus,
classical electron-electron and
exchange parts of the potential energy.
The strong orthogonality between the orbitals of 
the active ion A and those of the other ions S
is 
included  in H$_{eff}$ by means of the projection operator P$^S(i)$.
For closed shell ions, this operator can be written in the form\cite{Huz87} 
\begin{equation}
P^S(i) = \sum_{g \in S} \mid \phi_g^S> (-2 \epsilon_g^S ) <\phi_g^S \mid,
\end{equation}
where g runs over all occupied orbitals $\phi_g$ of the ion S
with orbital energies
$\epsilon_g^S$.
 
We now consider the explicit form of the effective potential 
$V_{eff}^S$\cite{Lua90}. 
For the  closed-shell ions  considered here 
the classical electrostatic part of this potential is given by
\begin{equation}
V_C^S(\vec r_1)= \int \rho^S (\vec r_2) r_{12}^{-1} d\tau_2
\end{equation}
where $\rho^S(\vec r)$ is the electron density of
ion S.  The exchange operator can be written 
as the nondiagonal
spectral resolution\cite{Huz87}
\begin{equation}
V_X^S(i)=-\sum_l \sum_{m=-l}^l \sum_{a,b} \mid alm,S> A(l,ab,S) < blm,S\mid,
\end{equation}
where $\mid alm,S>$ are products of spherical harmonics 
$Y_l^m$ and primitive radial
functions for the S ion, and $a$ and $b$ run over the Slater-type orbitals of
{\em l}-symmetry. $A(l,ab,S)$ are the elements of the matrix
\begin{equation}
A={\cal S}^{-1} {\cal K} {\cal S}^{-1}
\end{equation} 
and ${\cal S}$ and
${\cal K}$ are the
overlap and the exchange matrices for the S ion in the \{$\mid alm,S>$\}
basis.
More details can be found in
the original paper by Lua\~na and Pueyo\cite{Lua90}.

Finally the effective energy of the active ion $A$ is the sum of the electronic energy of 
eq.(4) plus the nuclear term 
\begin{equation}
E_{eff}^A (nucl)=-\sum_{S(\neq A)} Z^A v^S(\mid \vec R_A - \vec R_S \mid)
\end{equation}
where $v^S (\mid \vec R_A - \vec R_S \mid)$ is the electrostatic potential of the S group
at the nucleus of the active A ion. 
This potential contains nuclear and electronic contributions.

The practical method that we have used to apply this formulation to 
alkali-halide clusters is the
following one: for a given distribution of the ions
forming the cluster, we consider one of them as the active ion A (for
instance, a particular anion $I^{-}$), and solve the Self-Consistent-Field 
equations
for anion A
in the field of the remaining ions, which are considered frozen at this
stage. The solution
obtained is transferred to all the anions equivalent 
to anion A, that is, to the anions
which have  equivalent positions in the cluster. We take then a non
equivalent  $I^-$ anion (anion B) as the active ion and repeat the same process.
Evidently, since anions B are not equivalent to anions A, the energy
eigenvalues and wave functions of electrons in anions B can be 
different from those of anions A.
We continue this process 
in the same way until all the inequivalent anions have been
exhausted. 
The same procedure is then followed for the cations $Na^{+}$. The process
just described is a PI cycle. We iterate the PI cycles until
convergence in the total energy of the cluster is achieved.
Although for bulk NaI there are only two types of active ions ($Na^{+}$ and
$I^-$), this number is larger for  $(NaI)_n$ clusters. The reason is
that, in the cluster, $Na^{+}$ (and $I^-$) ions are in several different
environments.
Basis sets from Clementi and Roetti\cite{Cle74} for the ions (5s4p($Na^+$) 
and 11s9p5d($I^-$)) have been used in the calculations.
We have checked that this election is the best, between all the basis sets
reported by Clementi and Roetti, by carrying out some exploratory 
calculations for the
molecule and for $(NaI)_6$ with different basis sets.
More precisely, we have tested the use of the basis sets optimized
for $Na$, $Na^+$, $I$ and $I^-$. Between the four 
different possible combinations, the basis set for $Na^+,I^-$ led to the lowest 
total energies for the two clusters studied.
Inclusion of
diffuse functions
did not show any substantial improvement.
 
\section {Results.}

\subsection {Lowest energy structure and isomers.} 

In order to find the ground state structure we have to minimize the total 
energy $E$ of the cluster as a function of the positions {$\vec R_i$} of all the
ions. $E$ is a function of $3N - 6$ independent variables (N is the number of
ions) and minimizing that function in a multidimensional space in which we
must search for the absolute minimum among all the local minima is a tough
problem. A restricted search can be performed by starting with a number of
reasonable guesses for the geometry and allowing for the relaxation of those
initial geometries. As input geometries we
have considered cubic structures, rings and mixed structures, and for
specific numbers of molecules we have also considered more open structures
(like the wurtzite-type for $n=7$, or a truncated octahedron for $n=12$).
All those structures are suggested by previous works using 
pair-potentials\cite{Die85} or {\em ab initio} methods\cite{Ayu95}. The input
geometries have been optimized with respect to a few parameters. More
precisely, cuboid structures were relaxed with respect to a single parameter,
the first-neighbour cation-anion distance, whereas the ring structures were
optimized with respect to two or three parameters: the cation and anion
distances to the center of the ring, and the distance between rings.
However, for clusters with 7 molecules or less, we have performed, in
addition, a full relaxation of the structures obtained in the previous step.
To this end, we have used a simplex algorithm\cite{Nel65}, as given in
reference\cite{Wil91}. This is a simulated annealing recipe, which we have
used with the ``temperature'' set to zero, resulting in the conventional
downhill simplex method. The direct use of the simulated annealing technique
would be prohibitive for our calculational resources.

The results obtained for the geometries are shown in the figure 1.
The $Na^+$ and $I^-$ ions are represented by small and large
spheres, respectively.  The most stable structure (first isomer)
for each size is
shown on the left. The other structures are 
the low-lying isomers obtained (isomers several eV above the absolute 
minimum are not shown). In the case of
clusters with 3, 4 and 5 molecules we show two different views of the
second isomer. 
The total energy difference with respect to the absolute minimum is given for each 
isomer.
For n=2 we obtain a square as the ground state, 
for n=3
an hexagonal ring, for n=4 a cube,  
for n=5 a cube with a molecule attached to
an edge and for n=6 an hexagonal prism.
That is, we have found planar ground state structures for $n=2$ and $n=3$ and
three-dimensional structures for $n=4$ or larger. Planar ring-like isomers
exist for $n=4$ and $n=5$, quite low in energy in the second case. The cuboid
isomer for $n=6$ is also quite close in energy to the hexagonal prism. The
lowest energy structures for $n=2-6$ are, however, not perfect. The full
relaxation allowed by the calculation resulted in slightly distorted
geometries. Nevertheless for $n=5$ the distortion of the cube is large: the
cube edge facing the extra molecule enlarges its length. The structure for
$n=5$ can also be viewed as a distorted (non planar) hexagon with two
molecules attached, which can be justified in view of the structure of
$(NaI)_3$. The structure obtained for $n=7$ is a piece of the wurtzite crystal.

The same trend of competition between two different structural families
continues for $n$ larger than 7. $n=8$, $11$ and $14$ are cuboid
structures whereas $n=9$, $n=10$ and $n=15$ correspond to stacks of
hexagonal rings. Rather peculiar structures are obtained for $n=12$ and
$n=13$: a hollow cage for $n=12$ and a mixed structure (basket) for $n=13$. The
energy difference between the ground state and the cuboid isomers is so small
for $n=10$, $13$ and $15$, that a more flexible relaxation of the cuboid
isomer could perhaps change the order (we recall that the cuboid isomers have
been relaxed with respect to a single parameter). This is unlikely for
$n=9$, for which the energy difference reported in Fig. 1 is 0.14 eV. In fact,
we have performed additional calculations for the cuboid isomer in $(NaI)_9$
relaxing the energy with respect to more (six) parameters, but the hexagonal isomer
still remained as the lowest energy structure.

We proceed now to compare our results with
pair potential
calculations\cite{Die85}.
In the pair potential calculations two different models have been considered:
a rigid-ion model in which the energy is computed as a sum of
the electrostatic (Coulomb) and repulsive (Born-Mayer)
energies, and a polarizable-ion model in which each ion has
an induced dipole moment due to the fact that the closed
electronic shells are
polarized by the electric field of the others ions.
We find discrepancies with respect to the rigid-ion model for $n=4$, $5$, $8$,
$11$, $14$. In those cases the ground state predicted by that model
corresponds to our second isomer. Those structures contain hexagonal or
higher order rings. The same type of discrepancies are found compared to the
polarizable-ion model for $n=8$, $11$, $14$. Surprisingly, the
polarizable-ion model favors the defect-cuboid structure for $n=13$, which is
the second isomer in the PI calculations, only 0.02 eV above the ground
state. This suggest the possibility of the defect-cuboid becoming the
lowest energy structure in a PI calculation with a more flexible relaxation
of interatomic distances.

If we now compare the results of the PI model for $(NaCl)_n$\cite{Ayu95} and
$(NaI)_n$ we find a preference of the $NaI$ clusters for more open structures,
like the stacks of hexagonal rings for $n=6$, $9$, $10$ and $15$.
Our calculations show the same general trends in 
ground state structures shown in other {\em ab initio}
calculations when the cation size is quite smaller than the anion size.
For example, $(LiF)_n$ clusters studied in\cite{Och94} showed also a marked
preference towards ring-like (open) structures.
Thus, we can summarize these observations in the following way: if the ions
composing the cluster
have similar sizes, or if the cations are larger than the anions,
a preference for cubic structures against hexagonal prismatic forms is advanced
for those specific sizes in which both
structures are feasible, as in the case of clusters with
$6$, $9$ and $12$ 
molecules.
On the other hand, if anions are larger than cations, a preference for
hexagonal structures is advanced in the same cases. 
These hypothesis deserve further and more detailed 
studies on other ionic materials.

The $NaI$ crystal has the $NaCl$ structure\cite{Ash76}. The conclusion from the
study of small $(NaI)_n$ clusters ($n$ $\le$ $15$) is that only for some values
of $n$ the cluster can be considered as a small piece of the bulk lattice. But
there are other sizes, notably $n=6$, $7$, $9$, $10$, $12$, $13$, $15$, for
which the clusters, due to their very small sizes, prefer other structures.
The magic numbers observed in the mass spectra of large $NaI$ clusters\cite{Mar96}
(those clusters are, however, ionized and non-stoichiometric)
have been explained as cuboids with the bulk structure. We conclude that the
number of atoms in $NaI$ clusters with $15$ molecules or less is not yet enough
to produce bulk-like clusters.

\subsection {Inter-ionic distances}

In figure 2 we present the average interionic distance $d$ between
nearest neighbours $Na^+$ and $I^-$ ions for the
ground state geometry 
as a function of the cluster size.
$d$ presents an irregular behaviour
as a function of n, and this is due to the changes in the type of
structure. By joining the points corresponding
to  the same
type of structural family, either cubic or hexagonal, 
then the variation of $d$ is less
pronounced;
this is shown by the two dashed lines.
The
cubic clusters have an interatomic distance larger than the hexagonal ones.
$d$ tends to a saturation value ($\sim 3.4 \AA$) that is higher 
than the distance in the bulk ($d_{bulk}(NaI)=3.24 \AA$\cite{Ash76}).
This can perhaps be attributed to the lack of electron correlation in
the calculation.
The
incorporation of electron correlation tends to shorten the distances in 
ionic clusters, as other {\em ab initio} calculations\cite{Och94} have 
shown for $NaCl$ clusters.

\subsection {Binding energies}

The binding
energy per molecule of the cluster with respect to the separate
free ions, is given by:
\begin{equation}
E_{bind} = - \frac{1}{n} [E(cluster) - nE_0 (I^-) -
nE_0 (Na^+)] 
\end{equation}
where the energies of free $Na^+$ and $Cl^-$ ions 
are - 161.67692 a.u. and - 6918.06360 a.u,
respectively. $E_{bind}$
is shown as a function of $n$ in figure 3. The trend is an increase of the
binding energy with $n$.  But superposed to this general trend, 
especially stable clusters are predicted for
$n=(4)$, $6$, $9$, $12$. The stability of those clusters is reflected in local
peaks in $E_{bind}$ (or a pronounced change of slope for $n=4$). The magic
numbers agree with those predicted earlier for $(NaCl)_n$\cite{Ayu95}
using the PI model.
From the numbers given in figures 1 and 3 we notice that the energy differences
between isomers are very small compared to the total binding energies.
This observation is relevant because it indicates that the magic character of
some clusters (the peaks in fig.3) is associated to the specific number of 
molecules in those clusters and not to the particular structure of the lowest
isomer. In fact, $(NaCl)_n$ clusters also have magic numbers for $n= 6$, $9$,
$12$, although the lowest energy structures are cuboids (see fig.2 of
ref\cite{Ayu95}), that is, different from the lowest energy structures of
$(NaI)_n$. What makes special the number of molecules in the magic clusters
is that ``compact'' clusters can be built for those sizes.
We illustrate this with specific examples. The groun state of $(NaI)_5$ and
the hexagonal isomers 
of $(NaI)_7$ contain some low coordinated ions, in contrast
to $(NaI)_6$. Both the ground state and the cuboid isomer of $(NaI)_9$ are more
compact than the elongated forms of $(NaI)_8$ and $(NaI)_{10}$. Finally, the
hexagonal and cubic isomers of $(NaI)_{12}$ are again more compact than the
``defect'' forms of $(NaI)_{11}$ and $(NaI)_{13}$. This idea of stability
associated to compact clusters, which is evidently associated to the
optimization of the attractive part of the electrostatic energy, provides
justification for the interpretation of the periodic structure in the mass
spectrum of large $(NaI)_nNa^+$ clusters as revealing a cuboid pattern
growth\cite{Mar96}.

\subsection {Effects of the cluster size on the one-electron levels
and vertical ionization potentials.}

In figures 4 and 5, the eigenvalues 
of the 2p orbitals of $Na^+$ and the 5p orbital of $I^-$ ions are plotted
as a function of $n$. 
The results correspond to the
most stable structure for each cluster size.
Instead of a single eigenvalue we have a band of eigenvalues for each cluster
because the cations (or anions) occupy inequivalent positions in the cluster.
In other words, the 2p ($Na^+$) and 5p ($I^-$) eigenvalues are local atomic
properties, and as such these could determine the most reactive sites,
or the preferred adsorption sites.

The binding energy of the 2p level of the $Na^+$ ion, averaged over all
the ions in the cluster, decreases slowly with cluster size and the
average binding energy of the 5p level of the $I^-$ increases.
Of course, in the infinite crystal limit
all the cations are equivalent, so there will be only
one 2p eigenvalue for $Na^+$ (and a single
eigenvalue for the 5p level of $I^-$).

The clusters under study are made up of closed shells ions, whose
wave functions are strongly localized.
Thus, 
it can be assumed that
an electron is extracted from a specific localized orbital when the cluster
is ionized.  
This is the lowest bound $5p$ orbital, which generally corresponds to anions in
corner positions.
Using a Koopmans' like argument, we then identify the
vertical ionization potential of the cluster with the smallest
binding energy of a 5p electron. The dashed line in 
the figure 5 indicates the variation of the vertical 
ionization potential with the
cluster size.

The vertical ionization potential IP of a $(NaI)_n$
cluster in a pair potential model 
is a sum of two terms: the first is the energy necessary to remove an electron
from the isolated anion, i.e., the electronic affinity
EA of the neutral halogen atom, and the second is the electrostatic
interaction of the electron removed,
considered as a localized point charge,
with the other ions in the cluster, also considered unit point-like
charges, that is
\begin{equation}
IP=EA+ \sum_{j \ne i} \frac{q_i q_j} {r_{ij}}
\label{eq:ip}
\end {equation}
where $r_{ij}$ is the distance between ``i'' and ``j'' sites.  
The electrostatic term provides a strong stabilizing effect
(we note that the
EA of a free I atom is only 3.06 eV\cite{Hot75}).
On the other hand,
the short-range repulsive interaction among the
electron and the electrons of neighbouring ions has been neglected in 
equation (\ref{eq:ip}).
This interaction, which is
taken into account in the PI model, tends to lower the
ionization energy below the value given by equation (\ref{eq:ip}).

In order to study the different influence of the cluster field 
on cations and anions
we show in Table I the binding energy differences $\Delta(nl)$ between the
electronic levels ($nl$) of the ions in the cluster and in vacuum.
As a representative cluster we have chosen
the lowest isomer for $n=4$, relaxed with respect the nearest-neighbour distance.
The shifts $\Delta(1s)$, $\Delta(2s)$ and $\Delta(2p)$ 
are nearly the same for $Na^+$,
with differences in the meV range;  
this constant shift is the effect of the electrostatic term of
equation (\ref{eq:ip}).
However, the situation is different for the $I^-$ anions,
where the shifts are negative and larger 
for the outermost (5s and 5p) orbitals compared to the ``core'' orbitals.
This is indicative of the high polarizability of the $I^-$ anion.
The distinct stabilizing effects experienced by the outermost orbitals of the
$I^-$ anions compared to ``core'' orbitals is ascribed
to the quantum terms in the interaction energy,
namely, the exchange interaction and the wave function orthogonality.

\subsection {Effects of the cluster formation on the wave functions.
Partition of the binding energy}

The PI model assumes that the electronic density is a simple
superposition of weakly overlapping spherically symmetric
densities of the individual
ions.
%We proceed to study the changes
%in the densities upon formation of the cluster.
Upon formation of the cluster,
the $Na^+$ densities are practically equal to the free-ion
%Whereas the $Na^+$ densities are practically equal to the free-ion
densities, but a contraction of the $I^-$ density is found which is
%densities, a contraction of the $I^-$ density is found. This 
%deformation is mainly due to the 5p and 5s orbitals.
mainly due to the 5p and 5s orbitals. This is
in accordance with the general trends
for ionic solids\cite{Lua90}.
%In the figure 6 we plot the ``tails''
%of the 5s and 5p orbitals of $I^-$ in the $NaI$ 
%molecule.
%The orbital deformations are in accordance with the general trends
%for ionic solids\cite{Lua90}: they are appreciable only for the
%outermost anion orbitals,
%and the ion-cluster consistency achieved in the PI model
%becomes crucial to obtain a proper account of this effect. 

The orbital contraction of the outermost $I^-$ orbitals increases with the
cluster size, and is different for non equivalent anions. 
To get a more quantitative picture of this effect
we proceed to study its influence on the cluster energy.
The cluster binding energy of equation (13) can be written in a
convenient form to our purposes using the equations (1) and (3):
\begin {equation}
E_{bind} = -\frac{1}{n} \sum_{A=1}^N (E_{net}^A - E_O^A + \frac
{1}{2} E_{int}^A).
\end {equation}
By defining the deformation energy of the ion $A$ as:
\begin {equation}
E_{def}^A=E_{net}^A-E_O^A,
\end {equation}
and the binding energy of the same ion as:
\begin {equation}
E_{bind}^A= - (E_{def}^A + \frac{1}{2} E_{int}^A),
\end {equation}
we arrive to the following partition for the cluster binding energy:
\begin {equation}
E_{bind} = -\frac{1}{n} \sum_A (E_{def}^A + \frac{1}{2} E_
{int}^A) = \frac{1}{n} \sum_A E_{bind}^A.
\end {equation}
According to this equation, we can separate the cluster binding
energy into a sum of cationic and anionic contributions,
each one composed in turn of two terms:
one of them, $E_{def}^A$, accounts for the energy associated to the
deformation of the ionic wave functions and the other is half of the
interaction energy of that ion.
The first term, $E_{def}^A$, can never
be negative: we just have to recall that $E_{net}^A$ and 
$E_O^A$ are the expectation values of the intra-ionic part of the
hamiltonian
in states corresponding to the perturbed and the free ion,
respectively. Since the last one minimizes
that expectation value, we have $E_{net} ^A > E_O^A$. Thus, the
deformation energy is always an unfavourable contribution to the cohesion.
In contrast, the interaction energy term, is negative. 
The deformation and interaction parts of cationic and
anionic binding energies are given in Table II for clusters with $n=1$-$9$.
P indicates the specific type of site
when there are several
non equivalent positions.
For $n \le 7$, where we have made a
``full optimization'' of the structures, average values are quoted. We
notice the passive cationic response to
the cluster environment, the corresponding 
deformation energies being negligible. Instead, there are sizable
anionic deformation energies (up to 0.5 eV).  
The table distinguishes
differences between ions in vertex (V) and in edge (E) positions.
The deformation energies are smaller for anions in
vertex positions by 0.2 $eV$. 
This is related to the lower coordination
for ions in vertex positions.
However, in all cases, the anionic interaction energies 
are 
larger, in absolute value, than the corresponding deformation energies and
their sum is always negative.
Another interesting aspect is that we have at our disposal the
binding energies
of the ions at all nonequivalent positions in the cluster.

The binding energies in figure 3 grow quite fast with n,
but the limited size range studied here is not 
enough to estimate precisely the limiting value. That limit seems to be a
little smaller than the experimental binding energy in the bulk crystal (7.05
eV/molecule\cite{Ash76}). Our theoretical value for the molecule is also smaller than
the experimental one. Three factors can contribute to those errors.
In the 
first place, the geometries for $n>7$ have not been fully 
optimized.
Second, there is
an intrinsic variational restriction in the PI method, which is
associated to the neglect of covalent mixing between
anionic and cationic wave functions.
And third, we have neglected the 
coulomb electronic
correlation. We think that the main source of error is the last one,
in line with the arguments leading to the same conclusion in alkali halide
solids\cite{Lua90}.
Although it is sometimes stated that the
HF approximation would give good binding energies for
ionic compounds because the correlation errors in the free-ion and cluster wave
functions
tend to cancel out,
the orbital anionic deformation upon cluster formation casts
doubts on this assumption.

\section {Conclusions}

$(NaI)_n$ clusters up to $n=15$ have been studied using the {\em ab initio}
Perturbed Ion (PI) model.
Geometries have
been ``fully'' relaxed only for $n \le 7$, due to the increasing
computational time necessary to carry out these optimizations in larger 
clusters.
The isomeric study shows that these clusters are not yet fragments of the
bulk crystal lattice, although lattice fragments occur indeed for some values
of $n$.
Comparison to pair potential models shows some discrepancies.
Interionic distances do not show a smooth variation unless we restrict ourselves
to a fixed geometrical family. Though extrapolation to very large cluster sizes
is difficult,
we think that inclusion of 
correlation effects would be important in order to obtain quantitative
agreement with bulk quantities. Particularly stable (``magic'') 
clusters have been obtained for $n=(4)$, $6$, $9$ and $12$.

Cationic electron densities remain nearly undisturbed upon formation of the
cluster. On the contrary, anionic densities are contracted in
this process. The orbitals affected have been identified
as the outermost ones, mainly the 5p orbital of $I^-$. Nevertheless
the 5s orbital also experiments some deformation, and should
not be considered as a ``core'' orbital in a self-consistent-field
calculation. This effect has been analyzed by performing a study of the
energy eigenvalues.
In the PI model, the binding mechanism is a combination of the orbital
deformation and the quantum mechanical interaction 
between
each ion and the cluster, and 
the cluster binding energies have been analyzed from the point of view of this bonding
mechanism.

$\;$

$\;$

$\;$

$\;$

{\bf ACKNOWLEDGEMENTS}. This work has been supported by DGICYT (Grant
PB95-D720-C02-01). One of us (A. Aguado) acknowledges a predoctoral Grant from
Junta de Castilla y Le\'on.

\newpage

% Caption of tables
 {\bf Captions of tables}

$\;$

$\;$

{\bf Table I}. {Differences among the binding energies of the orbitals
   in $(NaI)_4$
    and the respective
   values for the ions in vacuum.  The shifts are given in eV.  }

$\;$

$\;$

{\bf Table II}. {Cation and anion deformation, interaction, and
binding energies for several cluster sizes (in $eV$). The
position P is given explicitly if there are different types of sites:  
edge (E), vertex (V).} 

\pagebreak

{\bf Captions of figures}

$\;$

$\;$

{\bf Figure 1}. {Isomeric geometries for $(NaI)_n$ clusters. The most stable
structure (first isomer) is shown on the left side. Total energy differences (in eV)
with respect to the most stable structure are
given for each isomer. $Na^+$, small spheres; $I^-$, large spheres.
For $n=3-5$, two different views of the second isomer are provided.

$\;$

$\;$

{\bf Figure 2}. Averaged nearest neighbour Na-I distance for the
structures of lowest total energy in Figure 1.
The two lines join cubic and hexagonal clusters, respectively. $n=7$, $12$,
$13$ correspond to peculiar structures.

$\;$

$\;$

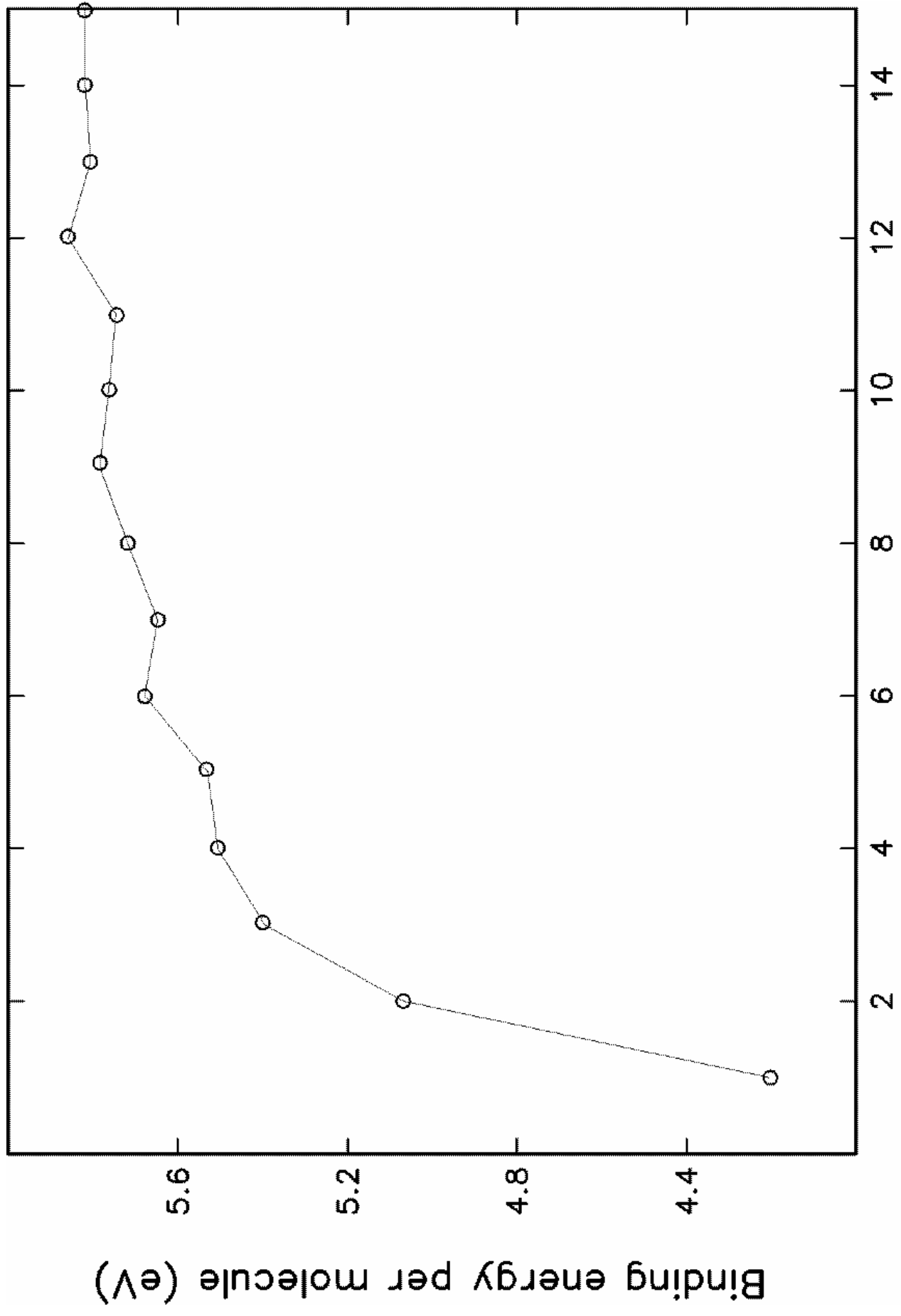
{\bf Figure 3}. Binding energy per molecule as a function of the cluster size.

$\;$

$\;$

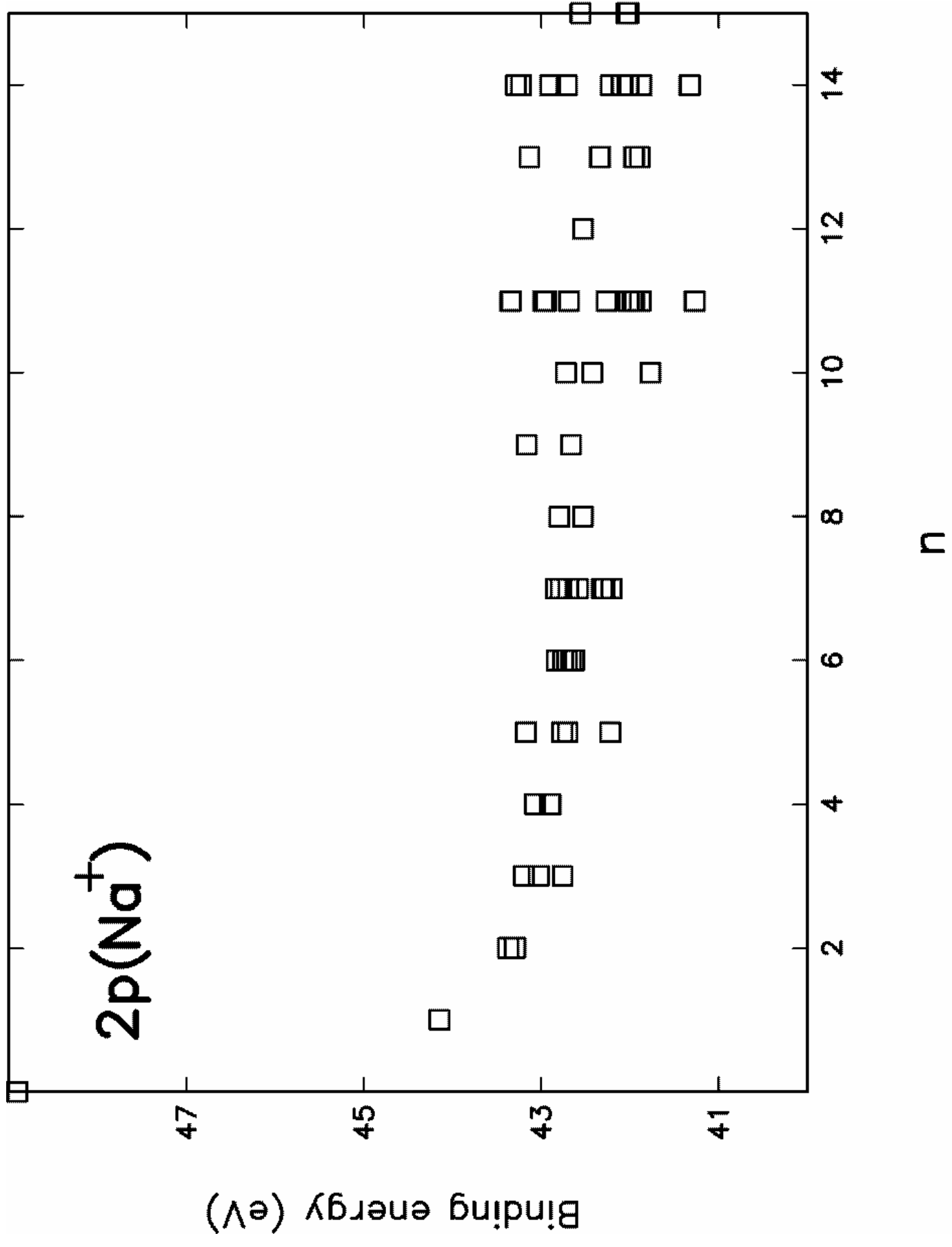
{\bf Figure 4}. Orbital energies (with opposite sign) of the 2p levels of $Na^+$
cations
as a function of the cluster size for the ground structures of
Figure 1.

$\;$

$\;$

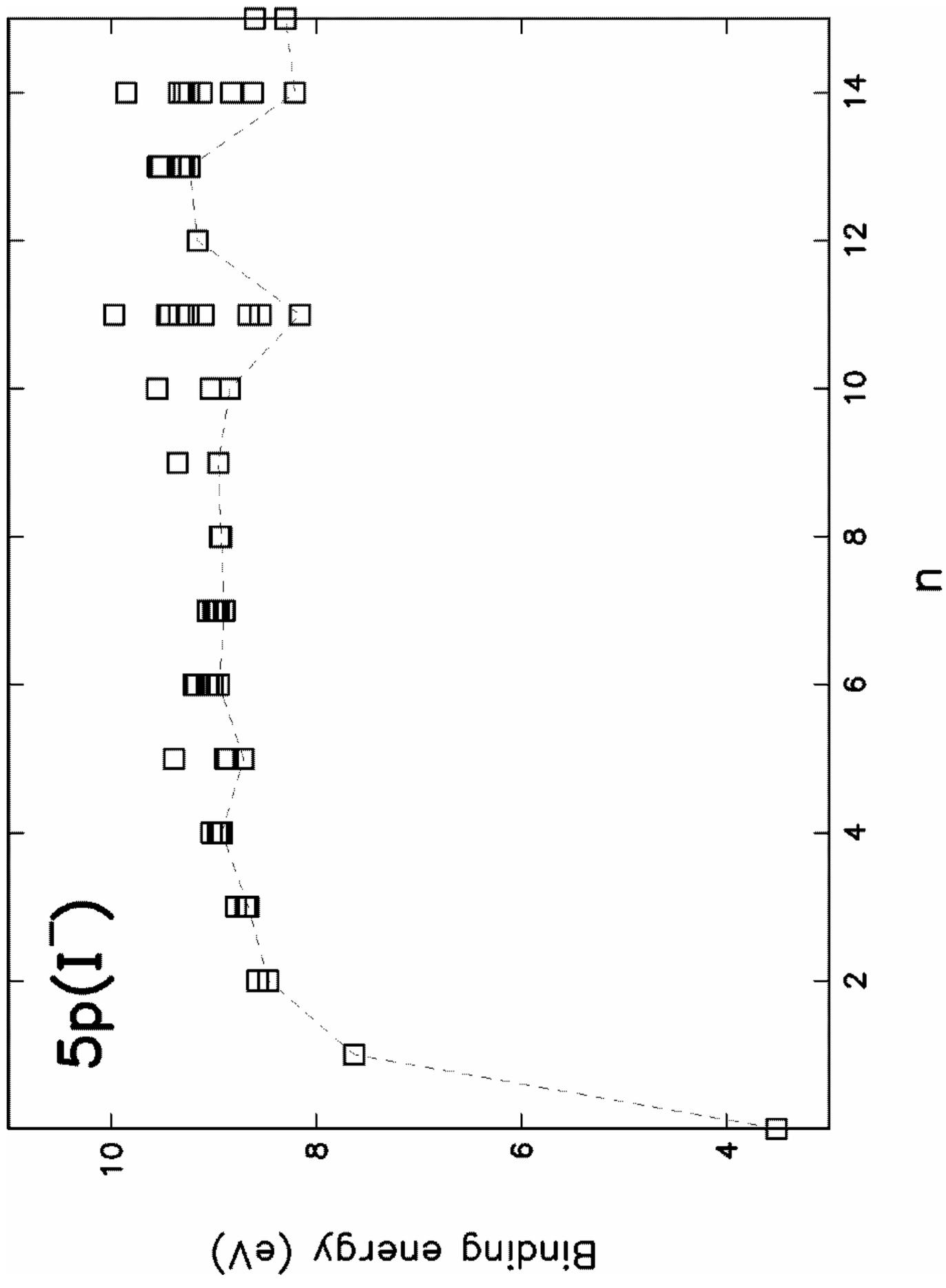
{\bf Figure 5}. Orbital energies (with opposite sign) of the 5p levels of $I^-$
anions
as a function of cluster size for the ground structures of $(NaI)_n$
clusters. The dashed line joins the vertical cluster ionization potentials.

%$\;$
%
%$\;$
%
%{\bf Figure 6}. Comparison of 5s and 5p orbitals of $I^-$ in vacuum (continuous
%lines) and
%for the $NaI$ molecule (dashed lines). 
%
\pagebreak

% Table I.

\begin {table}
\begin {center}

\bigskip

\begin {tabular} {||c|c|c|l|l|l||} \hline \hline
{$Na^+$} & (1s) & (2s) & & & \\
         & +6.239 & +6.237 & & & \\
         & & (2p) & & & \\
        & & +6.232 & & & \\
\hline
{$I^-$} & (1s) & (2s) & (3s) & (4s) & (5s) \\
         & -4.697 & -4.714 & -4.728 & -4.758 & -5.251 \\ & & (2p) & (3p) &
         (4p) & (5p) \\ & & -4.714 & -4.731 & -4.766 & -5.241 \\ & & & (3d) &
         (4d) & \\
        & & & -4.725 & -4.768 & \\
\hline
\end {tabular}

\end {center}
\label {table: endif}
\caption {Table I.}
\end {table}

\clearpage

\begin {table}
\begin {center}
\begin {tabular} {|c|c|c|c|c|c|c|c|c|}
\hline
\multicolumn {1}{|c|}{} &
\multicolumn {4}{c|}{$cations$} &
\multicolumn {4}{c|}{$anions$} \\
\hline
 n & E$_{def}$ & E$_{int}$ & E$_{bind}$ & P & E$_{def}$
 & E$_{int}$ & E$_{bind}$ & P \\
\hline
 1 & 0.0003 & - 4.8238 &  2.4116 & & 0.1301 & - 3.8487 &  1.7942 & \\  2 &
0.0002 & - 5.6798 &  2.8397 & & 0.2211 & - 5.0630 &  2.3104 & \\  3 & 0.0001
& - 6.0880 &  3.0439 & & 0.2582 & - 5.2000 &  2.3418 & \\  4 & 0.0002 & -
6.1880 &  3.0938 & & 0.3612 & - 5.7750 & 2.5263 & \\  5 & 0.0002 & - 6.3740
&  3.1868 & & 0.3217 & - 5.7310 &  2.5438 & \\  6 & 0.0005 & - 6.4750 & 
3.2370 & & 0.3620 & - 5.8300 &  2.5530 & \\
 7 & 0.0008 & - 6.4180 &  3.2082 & & 0.3684 & - 5.7100 &  2.4866 & \\
 8 & 0.0004 & - 6.3380 &  3.1686 & V & 0.2935 & - 5.7570 &  2.5850 & V \\
 & 0.0003 & - 6.5350 &  3.2672 & E  & 0.4861 & - 6.0110 &  2.5194 & E  \\
 9 & 0.0001 & - 5.9640 &  3.1915 & V & 0.3863 & - 4.6450 &  2.4652 & V \\
    & 0.0001 & - 6.5060 &  3.4721 & E  & 0.5969 & - 7.8230 &  2.5505 & E \\
\hline
\end {tabular}
\end {center}
\label {table:  ions}
\caption {Table II.}
\end {table}

\begin {figure} [htb]
\begin{center}
\unitlength=1mm
\begin{picture}(160.10,189.91)
\put(0,189.91){\special{em: graph FIG1A.PCX}}
\end{picture}
\end{center}
\end {figure}
\pagebreak
\begin {figure} [htb]
\begin{center}
\unitlength=1mm
\begin{picture}(160.10,189.91)
\put(0,189.91){\special{em: graph FIG1B.PCX}}
\end{picture}
\end{center}
\end {figure}
\pagebreak
\begin {figure} [htb]
\begin{center}
\unitlength=1mm
\begin{picture}(160.10,185.76)
\put(0,185.76){\special{em: graph FIG1C.PCX}}
\end{picture}
\end{center}
\end {figure}
\pagebreak
\begin {figure} [htb]
\begin{center}
\unitlength=1mm
\begin{picture}(160.10,190.08)
\put(0,190.08){\special{em: graph FIG1D.PCX}}
\end{picture}
\end{center}
\end {figure}
\pagebreak
\begin {figure} [htb]
\begin{center}
\unitlength=1mm
\begin{picture}(160.10,169.84)
\put(0,169.84){\special{em: graph FIG1E.PCX}}
\end{picture}
\end{center}
\end {figure}
\pagebreak
\begin {figure} [htb]
\begin{center}
\unitlength=1mm
\begin{picture}(150.11,122.09)
\put(0,122.09){\special{em: graph DIST.PCX}}
\end{picture}
\end{center}
\end {figure}
\pagebreak
\begin {figure} [htb]
\begin{center}
\unitlength=1mm
\begin{picture}(150.11,122.51)
\put(0,122.51){\special{em: graph ENBIND.PCX}}
\end{picture}
\end{center}
\end {figure}
\pagebreak
\begin {figure} [htb]
\begin{center}
\unitlength=1mm
\begin{picture}(150.11,125.56)
\put(0,125.56){\special{em: graph FIG4.PCX}}
\end{picture}
\end{center}
\end {figure}
\pagebreak
\begin {figure} [htb]
\begin{center}
\unitlength=1mm
\begin{picture}(150.11,122.00)
\put(0,122.00){\special{em: graph FIG5.PCX}}
\end{picture}
\end{center}
\end {figure}
\pagebreak
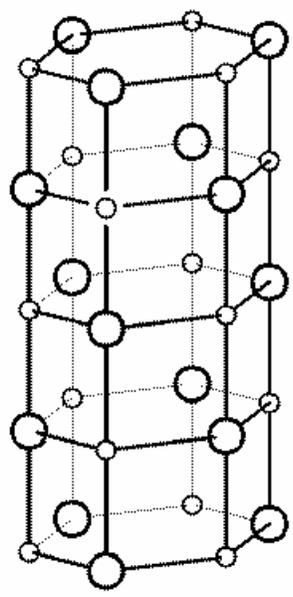
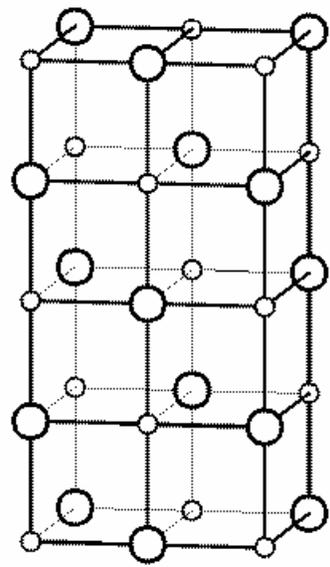
\end{document}